\documentclass[prd,aps,floats,twocolumn,nofootinbib]{revtex4-1}
\usepackage{slashed}
\usepackage{mathtools}
\usepackage{amsfonts}
\usepackage{amssymb}
\usepackage{epsfig}

\usepackage{bm}

\begin{document}

\newcommand{\m}[1]{\mathcal{#1}}
\newcommand{\nn}{\nonumber}
\newcommand{\ph}{\phantom}
\newcommand{\eps}{\epsilon}
\newcommand{\be}{\begin{equation}}
\newcommand{\ee}{\end{equation}}
\newcommand{\bea}{\begin{eqnarray}}
\newcommand{\eea}{\end{eqnarray}}
\newtheorem{conj}{Conjecture}

\newcommand{\plk}{\mathfrak{h}}


\title{Black holes and foliation-dependent physics
}
\date{}

\author{Jo\~{a}o Magueijo}
\email{j.magueijo@imperial.ac.uk}
\affiliation{Theoretical Physics Group, The Blackett Laboratory, Imperial College, Prince Consort Rd., London, SW7 2BZ, United Kingdom}

\begin{abstract}
In theories where physics depends on a global foliation of space-time, a black hole's horizon is surrounded by an ``eternity skin'': a pile-up of space-like leaves that in the far-out region cover all times from the start of collapse to future eternity. Any future foliation-dependent change in the laws of physics would be enacted in this region and affect the last stages of collapse towards black hole formation. We show how in some cases the black hole never forms but, rather, bounces into an explosive event. There is also a non-local transfer of energy between the asymptotic Universe and the formed black hole precursor, so that the back hole (if formed) or the exploding star (otherwise) will have a different mass from what was initial thrown in.  These last matters are generic to non-local theories and can be traced to the breakdown of the local Hamiltonian constraint.
\end{abstract}

\maketitle

\section{Introduction}

\begin{figure}[ht]
\begin{tabular}{c}
\includegraphics[trim=0cm 1cm 0cm 2cm, 
scale=0.4]{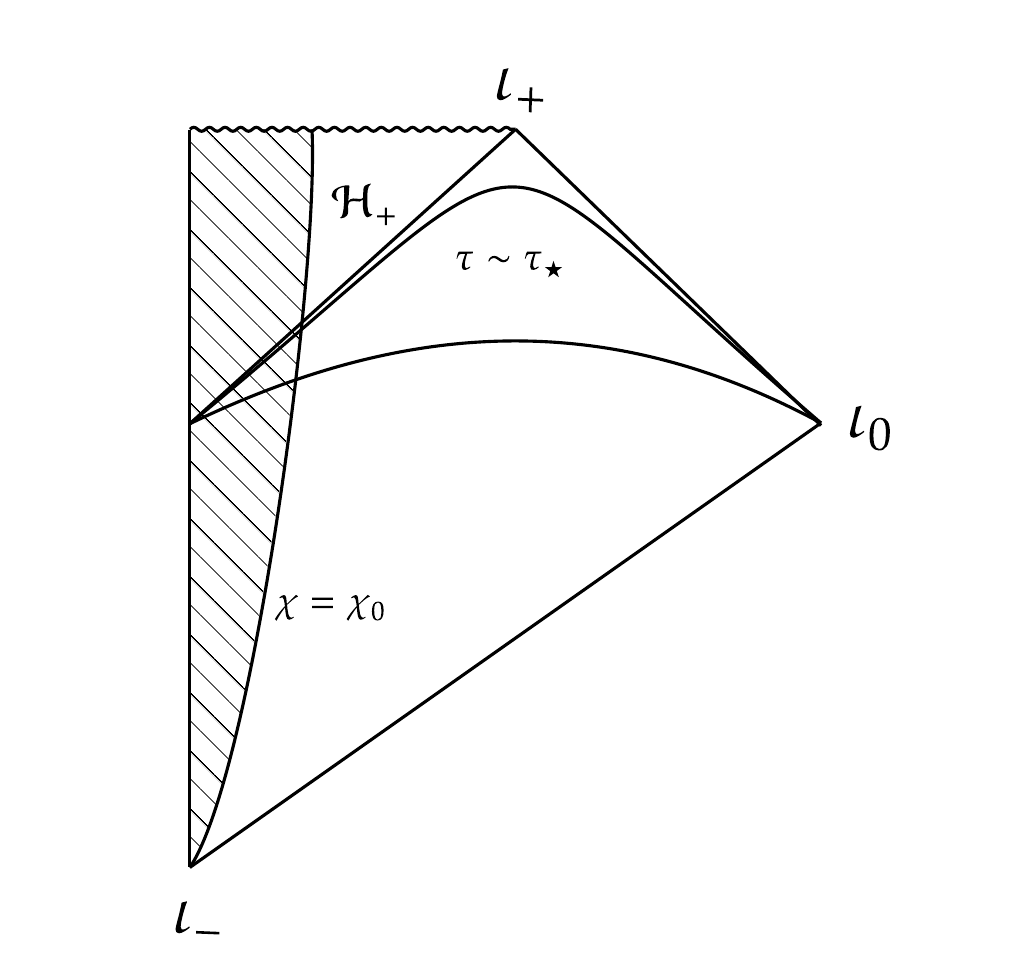}
\end{tabular}
\caption{The Penrose-Carter diagram of a standard dust ball collapse (with surface at $\chi=\chi_0$), with two leaves of the preferred foliation drawn. The upper leaf indicates a global event to the far future, at $\tau\sim \tau_\star$.  } \label{Penrose}
\end{figure}

The gravitational redshift effect around a black hole is a very interesting phenomenon. It implies that an observer in a far out region would have to wait an eternity to ``see'' a collapsing star plunge through the Schwarzchild radius, whereas an observer on the surface (or inside) such a star would experience an unremarkable finite time. More mathematically, this arises from the fact that the black hole horizon (${\cal H}^+$) is connected to the asymptotic time-like infinity ($\iota ^+$) in the conformal Penrose-Carter diagram (Fig.~\ref{Penrose})~\cite{HE}.  As a result, any asymptotic space-like surface (i.e., ending in $\iota^0$) to the future of the start of collapse  will become a space-like, near-null surface in the region just outside the black hole horizon, no matter how remote this future might be.

This fact potentially renders the black hole a remarkable futurological probe: the further into the future of the asymptotic outer Universe we want to probe, the closer to the black hole horizon we must go. If something dramatic were to happen {\it globally} in our future, this would be reflected in the environment near the horizon of any black hole, or in the final moments of collapse preceding their formation. Obviously, such global event is forbidden in theories where the laws of physics are purely local. But if physics were hypersurface dependent (namely if there were to be {\it future} evolution in the laws of physics in terms of a {\it global} time variable~\cite{evolution}), then this  would inevitably affect the local physics near the horizon of a black hole. This is the possibility to be investigated in this article.

In our pursuit we will be guided  by a few test tube examples. We consider only spherically symmetric black hole solutions, 
and focus mainly on the collapse of a dust ball within a Schwarzchild-like vacuole in a FRW Universe. This is a simple generalization of the Oppenheimer-Snyder model~\cite{OS,MTW,Stephani} and will be our prototype for a Machian set up, where a global outer Universe presides over local physics. In Section~\ref{etskin} we review the setting as it appears in General Relativity. Then, in Section~\ref{Machsetup} we show how it can relate to Mach's ideas, particularly as foliation dependent physics is introduced, allowing for direct action between the global Universe and local physics. We do this by defining global ``constants'' and their conjugate ``time variables'', allowing for variability in the laws of physics~\cite{evolution}. Interestingly, the local Hamiltonian constrain is lost during evolution (Section~\ref{HamStat}), even though it is still globally valid, and approximately true for the ``outer Universe'' in a Machian setting (Section~\ref{HamMach}). 

Having uncovered this important technicality, in Sections~\ref{OS-evol}--\ref{top-up}
we return to our toy model for black hole formation
in  theories where evolution takes place in the future. We find that a possible remarkable feature is that the collapse bounces into an explosive event and the black hole never forms. Whether the black hole forms or not, generically the mass of the final object is different than the initial mass, as a result of the energy conservation violation associated with evolution. These striking features can be interpreted as the effect of the Machian actions of the global Universe upon local physics.

\section{
The ``eternity skin'' in General Relativity}\label{etskin}
The presence of an ``eternity skin'' around a black hole is not controversial even in GR, although the question remains as to whether this could have physical effects (see~\cite{Rovelli-bounce}, for example). For a standard Schwarzchild black hole with metric:
\begin{equation}\label{Schw}
    ds^2=-\left(1-\frac{r_S}{r}\right)dt^2+\frac{dr^2}{1-\frac{r}{r_S}}+r^2d\Omega^2_2
\end{equation}
(where $r_S=2GM$ is the Schwarzchild radius, with $c=1$ in this Section) the asymptotically flat surfaces $\Sigma_t$ become outgoing near-null surfaces near the horizon, satisfying:
\begin{equation}
    t=r_\star = r+r_S\ln{\frac{r-r_S}{r_S}}.
\end{equation}
Since $t$ is also the proper time $t_\infty$ felt by these surfaces asymptotically far from the black hole, 
the ``eternity skin'' around the black hole horizon is therefore given by:
\begin{equation}
    \frac{r-r_S}{r_S}\approx e^{-\frac{\Delta t_\infty}{r_S}}.
\end{equation}
This expression relates how close to the horizon one must get to probe the asymptotic time lapse $\Delta t_\infty$ since black hole formation, as seen from infinity. The skin's spatial volume is:
\begin{equation}\label{skinvol}
    V(\Delta_t,\infty)=64\pi m^3 e^{-\frac{\Delta t}{2r_S}}.
\end{equation}
which is nothing but an expression of the exponential decay of the luminosity of a collapsing star, as seen from infinity. Hence, even without looking at the details of collapse, and assuming that the black hole has already formed, one can expect a mass top up coming from infinity in theories with matter creation to the future~\cite{evolution}, but we may expect this to be very small, simply by examining the volume factor involved (should the black hole form at all in such theories). We thus turn to the process of black hole formation.

Focusing specifically on a popular ``toy'' collapse model we take the Oppenheimer-Snyder (OS) homogeneous and isotropic collapsing dust ball~\cite{OS,MTW}. Inside the pressureless ball we have FRW with $k=1$ (although this is not needed; see~\cite{Stephani,Guth-OS}
for $k\neq 1$ examples):
\begin{equation}\label{FRW}
    ds^2=- d\tau^2+a^2(\tau)(d\chi^2+\sin^2\chi d\Omega^2_2)
\end{equation}
up to $\chi_0$ (which we assume $\chi_0\ll 1$). The solution is:
\begin{eqnarray}
     a(\eta)&=&\frac{a_m}{2}(1+\cos\eta)\\
     \tau(\eta)&=&\frac{a_m}{2}(\eta+\sin\eta)
\end{eqnarray}
where we have adjusted the integration constants so that the turnaround happens at $\eta=\tau=0$.
The matter constant of motion $m$ is:
\begin{equation}
    m=\rho a^3=\frac{3}{8\pi G }a_m
\end{equation}
obtained from the Friedmann equation evaluated at the $\dot a=0$ point
(with conventions for $a$ and angle $\chi$ so that $k=1$).
Outside the dust ball we have the Schwarzchild metric (\ref{Schw}), 
with geodesics given in parametric form by:
\begin{eqnarray}
    r&=&\frac{R_i}{2}(1+\cos\eta)\label{Rofeta}\\
    t&=&r_S\ln{\left| \frac{\sqrt{R_i/r_S-1}+\tan (\eta/2)}{\sqrt{R_i/r_S-1}-\tan (\eta/2)}\right|}\nn\\
    &&+r_S\sqrt{\frac{R_i}{r_S}-1}\left(\eta+\frac{R_i}{2r_S}(\eta +\sin\eta)\right),\label{tofeta}
\end{eqnarray}
for which the proper time is
\begin{equation}
    \tilde \tau=\sqrt{\frac{R_i^3}{4r_S}}(\eta+\sin\eta).
\end{equation}
For some $R_i$ (marking the boundary $r>R_i$ for which (\ref{Schw}) is valid) this external geodesic must coincide with the internal FRW geodesic $\chi=\chi_0$, for which the cosmological time, $\tau$, define by (\ref{FRW}), is also proper time. Hence $\tilde\tau=\tau$ for this geodesic, 
and we can relate the FRW proper time with $t$ by:
\begin{equation}\label{tauofeta}
     \tau=\sqrt{\frac{R_i^3}{4r_S}}(\eta+\sin\eta),
\end{equation}
together with (\ref{tofeta}). 
In addition, matching the extrinsic and intrinsic curvatures on both sides requires:
\begin{eqnarray}
    R_i&=&a_m\sin\chi_0\\
    r_S&=&a_m\sin^3\chi_0.
\end{eqnarray}
The above expressions define implicitly and parametrically a function $\tau(t)$, and its inverse. 
For some $\eta_\star$, as $\eta\rightarrow \eta_\star$ we have $r(\eta)\rightarrow r_S$ and $\tau\rightarrow \tau_\star$, both finite, but $t\rightarrow\infty$. This is the collapse equivalent of the eternity skin mentioned above for the vacuum solution. We will be examining the impact of evolution in the laws of physics upon this model.



\section{Foliation-dependent physics}\label{Machsetup}

Since its first  statement (Ref.~\cite{MachBook}, Chapter II, Section 6)
Mach's principle has developed a severe case of dissociative personality disorder. Of its ten-plus versions~\cite{MachReview} we will adopt the variant that states that the Universe as a whole affects (rather than fully determines) the local physics. This is the case in theories where the local physics depends on a foliation of space-time 
and on the global properties of the leaf where the local event lives. As an example we take theories mimicking the Henneaux-Teitelboim prescription~\cite{unimod} for implementing unimodular gravity~\cite{unimod1,unimod,UnimodLee1,alan,daughton,sorkin1,sorkin2}. If $S_0$ is a base action depending on ``parameters''  $\bm\alpha$ (``constants'', as it were), one adds a term to the action:
\begin{eqnarray}
      S&=& \int d^4x\, (\partial_\mu {\bm \alpha})\cdot {\cal T}_{\bm\alpha}^\mu +S_0,
\end{eqnarray}
where ${\cal T}_{\bm\alpha}$ is a density. This density is then used to produce a foliation-dependent time variable~\cite{unimod,Bombelli,sorkin1,sorkin2,UnimodLee2,JoaoLetter,JoaoPaper} as follows.
Performing a 3+1 split, and 
solving for the Lagrange mutiplier ${\cal T}^i$, we first obtain spatial constancy of $\bm \alpha$ on the leaves $\Sigma_t$.  Retaining the gauge invariant zero-mode of ${\cal T}^0$ (see~\cite{unimod} for details) we then define a time variable:
\begin{equation}
    {\bm T}_{\bm\alpha}(t)=\frac{1}{V_c}\int_{\Sigma_t} d^3x\, {\cal T}^0_{\bm\alpha}.
\end{equation}
This is divided by the spatial coordinate volume $V_c$, so that time is intensive~\cite{JoaoLetter,JoaoPaper}. For example, in the unimodular case, for which the target $\alpha$ is $\rho_\Lambda$ (the vacuum energy), the time $T_\Lambda$ is 4-volume per unit of coordinate (comoving, in the case of FRW) 3-volume.

Under this prescription, the constants $\bm\alpha$ and their times are conjugate variables that depend only on $t$, since the action reads:
\begin{equation}
    S=V_c\int dt\, \dot {\bm\alpha}\cdot {\bm T}_{\bm\alpha} +S_0,
\end{equation}
implying:
\begin{equation}
    \{{\bm\alpha}, {\bm T}_{\bm\alpha}\}=\frac{1}{V_c}.
\end{equation}
Because they only depend on $t$, they are global variables, defined leaf by leaf.
The set up fixes a preferred foliation $\Sigma_t$, and this is particularly pertinent if we impose ${\bm T}_{\bm\alpha}$ dependence upon $S_0$, a situation described as ``evolution in the laws of physics'' in~\cite{evolution}.
For example, $S_0$ may contain {\it other} parameters $\bm\beta$ which are functions of ${\bm T}_{\bm\alpha}$, according to functions $\bm\beta({\bm T}_{\bm\alpha})$, akin to evolution potentials~\cite{evolution}.
Then, the local physics (dependent on parameters $\bm\alpha$ and $\bm\beta$) is affected by the global variables defined on the leaves $\Sigma_t$. We stress that this construction 
may select a foliation, but it certainly is an improvement upon a  {\it coordinate} time-dependence, since the times used are physical, and so the preferred foliation is fixed by physical matter or geometrical entities.

\begin{figure}[ht]
\begin{tabular}{c}
\includegraphics[clip,
trim=0cm 4cm 4cm 0cm
scale=0.7]{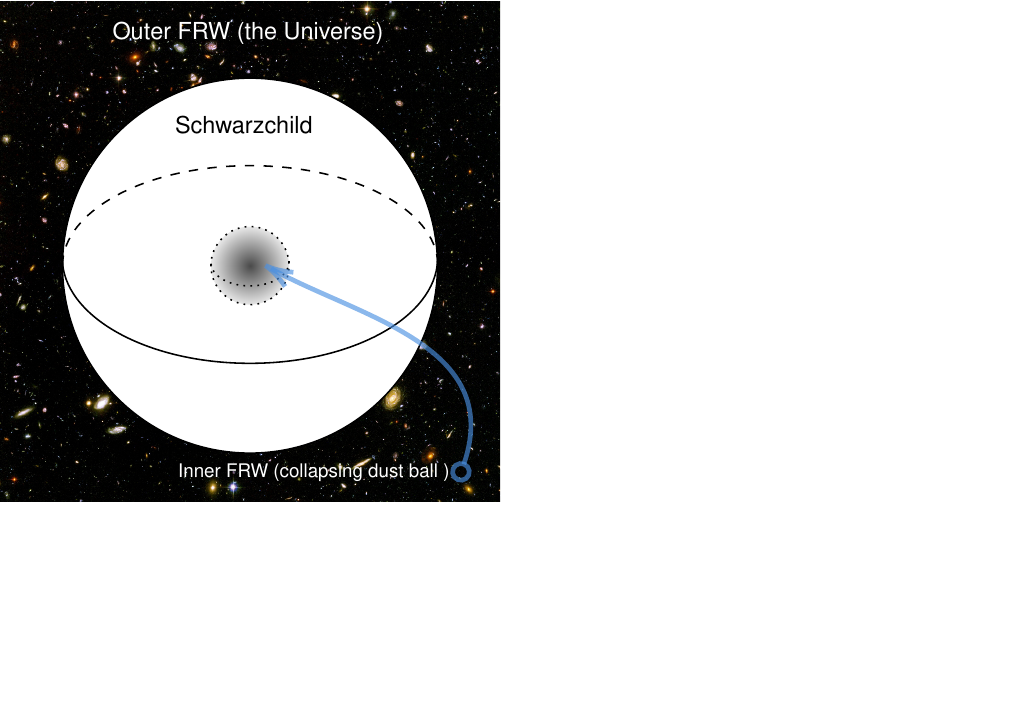}
\end{tabular}
\caption{Our set up is a hybrid between the Oppenheimer-Snyder model and a vacuole. Beside the inner collapsing FRW dust model (playing ``Newton's bucket''), we consider an outer FRW model (Mach's ``fixed stars'') linked to the Schwarzchild solution in its far out region. In the usual theory the outer model does not affect the inner collapse, but in the presence of global variables it does, pointing to a breakdown of Birkhoff's theorem. } \label{Machset}
\end{figure}

Mach's concept is predicated on the distinction between ``local'' (e.g. a rotating bucket) and ``Universe'' (the ``fixed stars''). Obviously the Universe is made of a collection of locals, but it is by coarse-graining their details and letting them gang up as a whole that the ``Universe'' is defined. In Mach's vision we let the Universe affect a ``local'' detail we have decided to focus on, be it a rotating bucket or the formation of a black hole. In this spirit, we consider the OS model described in the previous Section, embedded in an outer FRW dust Universe, with metric:
\begin{equation}\label{outFRW}
    ds^2=-d\bar t^2+\bar a^2 \left(\frac{d\rho^2}{1-\bar k\rho^2}+\rho^2d\Omega_2^2\right),
\end{equation}
providing the Universe. 
We could for example consider a spherical collapse starting in the outer FRW (e.g.~\cite{Stephani}). This would create a vacuole, with an OS model in the middle. The internal collapsing FRW patch of the OS model is then connected to the asymptotic outer FRW via the intermediate Schwarzchild solution (see Fig.~\ref{Machset}).

Assuming that the vacuole is sufficiently large, the outer FRW cosmological proper time $\bar t$ is the same as Schwarzchild's $t$:
\begin{equation}\label{ttildet}
    \bar t\approx t.
\end{equation}
Hence, the outer cosmological proper time, $t$, and that of the OS inner FRW model, $\tau$, are implicitly related by (\ref{tofeta}) and (\ref{tauofeta}) (with $\tau=\tilde\tau$ as discussed). There is a $\tau_\star$ such that as $\tau\rightarrow\tau_\star$ we have $t\rightarrow\infty$ (and $R\rightarrow r_S$). This formalizes the eternity peel in this context. 
All we need now is to find the dynamics for a theory with evolution within this setup. However, a major difference is found with respect to minisuperspace reductions, as studied in~\cite{evolution}, as we now explain. 

\section{Status of the  Hamiltonian constraint}\label{HamStat}

The crucial novelty is that 
if there is
${\bm T}_{\bm\alpha}$ dependence in $S_0$ (e.g., via $\bm\beta({\bm T}_{\bm\alpha})$), then we lose the local, but not the global Hamiltonian constraint. This is because the theory is only invariant under lapses and shifts of the form:
\begin{eqnarray}
    N&=&N(t)\\
    N^i&=&N^i(t,x^j)
\end{eqnarray}
that is, $N$ is not allowed to depend on $x^i$. These are the transformations that preserve the foliation $\Sigma_t$, and they are the symmetries of any other foliation-dependent theory, such as Horava-Lifshitz theory, cf.~\cite{HL}. By varying with respect to $N$ we obtain the global Hamiltonian constraint:
\begin{equation}\label{Hamconst}
    H=N(t)\int d^3 x\, {\cal H}(x)= 0,
\end{equation}
yet, there is no local Hamiltonian constraint, 
${\cal H}(x)=0$.

Another way to see that under evolution ${\cal H}=0$
could not be true is to examine its associated secondary constraint
$\dot {\cal H}=0$. This cannot be enforced, as we now show. 
It is true that $\bm\beta({\bm T}_{\bm\alpha})$ evolution 
avoids a preferred foliation defined directly in terms of coordinate time, and so the local Hamiltonian is still independent of coordinate $t$. However, it does depend on global (``Machian'')  variables, and this spoils the usual argument leading to $\dot {\cal H}=0$ even if ${\cal H}$ does not depend on $t$. Indeed, writing
${\cal H}(x)={\cal H}(q(x),p(x);{\bm\alpha}, {\bm T}_{\bm\alpha})$, where $q(x)$ and $p(x)$ are generic local degrees of freedom, we can expand (suppressing the $x$ labels for clarity):
\begin{eqnarray}
   \dot {\cal H} &=&\frac{\partial {\cal H}}{\partial t} +
   \frac{\partial {\cal H}}{\partial q}\dot q +\frac{\partial {\cal H}}{\partial p}\dot p+ 
   \frac{\partial {\cal H}}{\partial{\bm \alpha}}\dot {\bm \alpha}+
   \frac{\partial {\cal H}}{\partial{\bm T}_{\bm \alpha}}\dot {\bm T}_{\bm \alpha}\nn\\
   &=&  \frac{\partial {\cal H}}{\partial{\bm \alpha}}\dot {\bm \alpha}+
   \frac{\partial {\cal H}}{\partial{\bm T}_{\bm \alpha}}\dot {\bm T}_{\bm \alpha}
   \label{dotcalH}
\end{eqnarray}
where we have used Hamilton's equations for the local variables to cancel the second and third terms. However, the Hamilton equations for the global variables are:
\begin{eqnarray}
\dot {\bm \alpha}&=&\{\bm\alpha,H\}=\frac{1}{V_c}\frac{\partial H}{\partial {\bm T}_{\bm \alpha}}=\frac{N(t)}{V_c}\int d^3x\, \frac{\partial {\cal H}}{\partial {\bm T}_{\bm \alpha}}\label{eqalpha}\\
 \dot {\bm T}_{\bm \alpha}   &=&\{{\bm T}_{\bm\alpha},H\}=-\frac{1}{V_c}\frac{\partial H}{\partial {\bm \alpha}}=-
 \frac{N(t)}{V_c}\int d^3x\, \frac{\partial {\cal H}}{\partial {\bm \alpha}}\quad \label{eqTalpha}
\end{eqnarray}
i.e., they employ $H$, not $\cal H$. 
This spoils the equivalent cancellation of the remaining terms, with: 
\begin{equation}\label{dotHloc}
  \dot {\cal H}=\frac{1}{V_c}\left(
  \frac{\partial {\cal H}}{\partial{\bm \alpha}}\frac{\partial H}{\partial {\bm T}_{\bm \alpha}}-
   \frac{\partial {\cal H}}{\partial{\bm T}_{\bm \alpha}}\frac{\partial H}{\partial {\bm \alpha}}\right)
  \neq 0  .
\end{equation}
A more compact derivation leading to the same result, appeals to the Poisson bracket:
\begin{eqnarray}
   \dot {\cal H} &=&\frac{\partial {\cal H}}{\partial t} +  \{ {\cal H},H \}=\{ {\cal H},H \}_{NL}
\end{eqnarray}
where we denoted by $\{ f,g\}_{NL}$ the terms in the Poisson bracket involving the non-local variables. This alternative derivation shows how $\dot {\cal H}\neq0$  can be traced to:
\begin{equation}
    \{ {\cal H}(x),{\cal H}(x')\}\neq 0
\end{equation}
(with $x\neq x'$) due to the dependence of ${\cal H}$ on $\bm \alpha$ and  ${\bm T}_{\bm \alpha}$, since  we can also write:
\begin{equation}
    \dot {\cal H}(x)=\{{\cal H}(x),H\}=N\int d^3 x'\, \{{\cal H}(x),{\cal H}(x')\}.
\end{equation}
Whichever expression we use the effect disappears if integrated over the whole space:
\begin{equation}
    \int d^3x\, \dot {\cal H}(x)=0, 
\end{equation}
so that (\ref{Hamconst}) is consistent. 

Hence, we find something that is absent in the usual FRW  or minisuperspace (MSS) set up. In that case there is significant redundancy in obtaining the dynamics~\cite{evolution}. Recall that in the standard theory (with no evolution) one can derive energy-momentum conservation either from the gravitational equations and Bianchi identities (which amount to the secondary constraint $\dot H=0$), or directly from the matter equations of motion.
The same redundancy applies to theories with time-dependence (evolution) when reduced to MSS~\cite{evolution}, with both methods leading to the same source terms in the conservation equation. This redundancy is lost  if we have evolution {\it and} go beyond homogeneity and isotropy, and focus on local equations, because we lose the Hamiltonian constraint, and so the Bianchi identity. 


\section{Hamiltonian constraint in a  Machian setting}\label{HamMach}

The general statements made in the previous Section, take a special form in a Machian setting, where we can split the system into a presiding Universe (here assumed homogeneous) with volume $V_\infty$ and Hamiltonian 
\begin{equation}
 H_\infty=N(t) V_\infty {\cal H}_\infty(t)   
\end{equation}
and a local system with volume $V_L\ll V_\infty$ and Hamiltonian
\begin{equation}
    H_L=N(t)\int_L d^3 x\, {\cal H}_L(x).
\end{equation}
Then, 
Eqs. (\ref{eqalpha}) and (\ref{eqTalpha}) become approximately:
\begin{eqnarray}
\dot {\bm \alpha}&=&\frac{N(t)}{V_c}\left (V_\infty \frac{\partial {\cal H}_\infty}{\partial {\bm T}_{\bm \alpha}}+\int_{L} d^3x\, \frac{\partial {\cal H}_L}{\partial {\bm T}_{\bm \alpha}}
\right)\approx N(t) \frac{\partial {\cal H}_\infty}{\partial {\bm T}_{\bm \alpha}}\nn
\\
 \dot {\bm T}_{\bm \alpha}   &=&-
\frac{N(t)}{V_c}\left (V_\infty \frac{\partial {\cal H}_\infty}{\partial {\bm \alpha}}+\int_{L} d^3x\, \frac{\partial {\cal H}_L}{\partial {\bm \alpha}}
\right)\nn\\
&\approx& -N(t) \frac{\partial {\cal H}_\infty}{\partial {\bm \alpha}},\label{GlobalApprox}
\end{eqnarray}
that is, when determining the global variables 
in a Machian setting, the presiding Universe dominates due to its volume.

A direct implication is that, even though there is no local Hamiltonian constraint (${\cal H}_L\neq 0$) under evolution, there is still an approximate Hamiltonian constraint for the Machian Universe, ${\cal H}_\infty=0$. The obstruction
(\ref{dotcalH}) is removed, since inserting (\ref{GlobalApprox}) leads to:
\begin{eqnarray}
   \dot {\cal H}_\infty &=&
   \frac{\partial {\cal H}_\infty}{\partial{\bm \alpha}}\dot {\bm \alpha}+
   \frac{\partial {\cal H}_\infty}{\partial{\bm T}_{\bm \alpha}}\dot {\bm T}_{\bm \alpha}\nn\\
   &\approx &N\left(\frac{\partial {\cal H}_\infty}{\partial{\bm \alpha}}\frac{\partial {\cal H}_\infty}{\partial {\bm T}_{\bm \alpha}}-
   \frac{\partial {\cal H}_\infty}{\partial{\bm T}_{\bm \alpha}}\frac{\partial {\cal H}_\infty}{\partial {\bm \alpha}}\right) =0,
\end{eqnarray}
and indeed
varying the action with respect to $N$ we find: 
\begin{equation}
    0=H=V_L{\cal H}_L+V_\infty{\cal H}_\infty\approx V_\infty{\cal H}_\infty.
\end{equation}
This makes sense, since it is the result found under the assumption of homogeneity~\cite{evolution}, 
and this assumption is nothing but the result of coarse-graining detail, and focusing on what wins by volume. 

We add a few comments on this interesting result. Firstly, in a Machian setting, the global variables are an essentially {\it external} input into the local physics. They affect the local physics, but their dynamics are not dependent upon it. Instead their dynamics can be inferred from the minisuperspace solutions, independently of the local setting one considers. We can plug any of the solutions found in~\cite{evolution} into the local physics, independently of the latter details.

Secondly, the Machian ``outer'' Universe does not even need to be our observable Universe; indeed, if there is a cosmological horizon effect, it is not. Rather, it is the whole Universe contained in each leaf $\Sigma_t$. By Occam's razor we can assume that whatever is outside our cosmological horizon, and dominates the global variables' equations, has the same average properties we observe in a smoothed version of our horizon, but this is not necessary. This possibility has interesting implication for our local FRW Universe, as we will investigate in~\cite{evol-CDM}.

Finally, {\it if} we identify the notorious problem of time in quantum gravity~\cite{time-prob,time-prob1} with the Hamiltonian constraint in General Relativity, then we can draw the following conclusions in the context of this theory. 
The less evolution there is (i.e. the more constant the constants are), the more we have a problem of time (i.e. a Hamiltonian constraint). But also should there be evolution, the bigger the chunk of the Universe we consider and coarse grain, then the bigger the problem of time that is left unaffected by evolution. Still, even under the tiniest amounts of evolution, the {\it local} scales will always see a time, because the Hamiltonian constraint is violated for them. The problem of time is then a problem for both theories without evolution, and in theories with evolution for the global coarse-grained Universe. Each local region living therein fails to see what is the problem.

\section{foliation-dependent Oppenheimer-Snyder collapse}\label{OS-evol}
We can now match the setting described at the end of Section~\ref{Machsetup} (an OS model embedded in an  outer FRW) into our framework. We take for local system (${\cal H}_L$) the inner collapsing dust FRW with metric (\ref{FRW}). For Machian Universe we take the outer FRW metric (\ref{outFRW}), providing ${\cal H}_\infty$ and  responsible for the dynamics of the global variables, $\bm\alpha$, $\bm\beta$ and ${\bm T}_{\bm\alpha}$. Connecting the two we have a  Schwarzchild region.  

Assume first that the functions $\bm\beta({\bm T}_{\bm\alpha})$ are either pulses (delta functions of the form $\bm\beta=\delta_{\bm\beta} \,\delta(\bm T_{\bm\alpha}-\bm T_{\bm\alpha \star})$) or steps (Heaviside functions of the form  ${\bm\beta}=\Delta {\bm \beta} \, H(\bm T_{\bm\alpha}-\bm T_{\bm\alpha \star})$) in the far future. Hence, at least up to the time $\bm T_{\bm\alpha \star}$, we can assume the standard relation between the Schwarzchild coordinate $t$ (with $t\approx \tilde t$, the 
asymptotic FRW proper time, cf. (\ref{ttildet})) and the inner FRW proper time $\tau$, as given implicitly by (\ref{tofeta}) and (\ref{tauofeta}).
Generally ${\cal H}_\infty$ determines $T_{\bm\alpha}(t)$ via (\ref{GlobalApprox}). Hence when evaluating the effect upon the OS inner  FRW we 
must compute the chain of composite functions: 
\begin{equation}
  \bm\beta(\tau)=\bm\beta(\bm T_{\bm\alpha}(t(\tau))).
\end{equation}
There is a $\tau_\star$ such that as $\tau\rightarrow\tau_\star$ we have $t\rightarrow\infty$ (and $R\rightarrow r_S$). Hence we do not even need to strictly assume that $\bm\beta({\bm T}_{\bm\alpha})$ are pulses or steps, since any function varying only over a finite duration in the future will be mapped into a pulse or a step (or a combination) in  $\bm\beta(\tau)$ at $\tau=\tau_\star$.

To fix ideas, we focus on models extracted from a theory where for clock generators we have:
\begin{eqnarray}
 {\bm \alpha}&=&\left(\rho_\Lambda, \frac{3 c_P^2}{8\pi G} \right),\label{alpha}
\end{eqnarray}
where $\rho_\Lambda$ is the vacuum energy, $c_P$ is the speed of light as it appears in the gravitational commutation relations~\cite{EllisVSL,evolution,JoaoPaper,JoaoLetter}, and $G$ is the gravitational constant~\cite{vikman,vikman1,vikaxion,JoaoPaper,JoaoLetter}. This is similar to sequestration~\cite{padilla,pad,pad1,vikman,vikman1,vikaxion,lomb} and for base action:
\be\label{S0def}
S_0=\int d^4 x\, \sqrt{-g}\left[\frac{\alpha_2}{6}R+{\cal L}_M -\rho_\Lambda\right],
\ee
(or its Einstein-Cartan variation~\cite{evolution}, as we will assume here) the associated times are:
\begin{eqnarray}
T_1(\Sigma_t)\equiv T_\Lambda(\Sigma_t)&=&
-\int^{\Sigma_t}_{\Sigma_0} d^4 x \sqrt{-g}\label{TLambda}
\\
    T_2(\Sigma_t)\equiv T_R(\Sigma_t)&=&\frac{1}{6}\int^{\Sigma_t}_{\Sigma_0} d^4 x \sqrt{-g} R\label{TRicci}.
\end{eqnarray}
that is, minus the 4-volume between leaf $\Sigma_t$ and a conventional ``zero-time'' $\Sigma_0$ leaf to its past, or unimodular time~\cite{unimod}, and 
its Ricci weighted counterpart. 
(We will often omit the $\Sigma$ in our expressions.)
For simplicity we will only consider the speed of gravity,
$c_g^2$ as a potential $\bm\beta$ (see~\cite{AM,VSLreview} for some background). 
The matter action is that of dust for the inner FRW or a generic perfect fluid (which may be dust) for the outer FRW, as described in~\cite{brown,gielen,gielen1,evolution}.

Reducing the Einstein-Cartan action to our setting, we have two FRW situations, one for the collapsing dust ball, another for the presiding Universe: 
\begin{eqnarray}
    S_L&=&V_L\int dt\, [\alpha_2 \dot b a^2 +\dot m T_m- N{\cal H}_L]\label{SL}\\
    {\cal H}_L&=&a\left(-\alpha_2 (b^2+kc_g^2)+\frac{m}{a}\right)\label{HL}\\
    S_\infty&=&V_\infty\int dt\, [\alpha_2 \dot {\Bar b} \Bar a^2 +\dot {\Bar m} \Bar T_m- N{\cal H}_\infty]\label{SU}\\
    {\cal H}_\infty&=&\Bar a\left(-\alpha_2 (\Bar b^2+\Bar kc_g^2)+\frac{\Bar m}{\Bar a^{1+3w}}\right).
\end{eqnarray}
Here $a$ is the expansion factor, $b$ the FRW connection variable (which in the standard theory is $\dot a$ on-shell, that is the comoving inverse Hubble length), and the reduced fluids have been represented following~\cite{brown,gielen,gielen1,evolution}, with $w$ the equation of state and $m$ their conserved quantity. We use a bar for the outer variables and no bar for the inner variables. 
This defines the Poisson brackets for all variables, as well as the Hamiltonian. The $N$ is common to the two systems.

\section{Bouncing from collapse to explosion}
Let us consider the evolution potential $c^2_g=c^2_g(T_R)$ to illustrate the general points made in Sections~\ref{HamStat} and~\ref{HamMach}. 
The fact that we lose the Hamiltonian constraint ${\cal H}_L=0$ does not mean that this is not a Hamiltonian system (most Hamiltonian systems do not have a Hamiltonian constraint after all). The usual framework applies, except that care must be taken not to use standard manipulations which implicitly use the Hamiltonian constraint. The Hamilton's equations lead to~\cite{evolution}:
\begin{eqnarray}
     \dot a+\frac{\dot \alpha_2}{2\alpha_2}a&=& Nb\label{dotaG}\\
    \dot b&=&-\frac{N}{2 a} (b^2+kc_g^2) 
    \label{dotbG}\\
    \dot \rho+3\frac{\dot a}{a}\rho&=&0.
\end{eqnarray}
The first equation contains a correction typical of theories with varying Planck Mass in the Einstein-Cartan framework~\cite{evolution,flanagan}. 
The second equation is the Raychaudhuri equation stripped of its standard simplification using the Friedman equations.
The third equation arises directly from 
\bea
\dot m&=&\{m,H\}=0,
\eea
rather than combining the first two equations with the Hamiltonian constraint, as is usual both in standard theory and under evolution in MSS~\cite{evolution} (see comments on redundancy made at the end of Section~\ref{HamStat}).

For the asymptotic FRW (assuming $\bar w=0$, but this is not necessary\footnote{Even with $\bar w=0$, the cosmic dust, $\Bar\rho$, does not need to be the same matter specie as the collapsing dust ball, $\rho$. For example one could be ``dark matter'', the other ``baryonic matter''.}) we have the usual~\cite{evolution}:
\begin{eqnarray}
\Bar  b^2 + c_g^2 \Bar k &\approx& \frac{\Bar \rho {\Bar a}^2}{\alpha_2}\\
     \dot {\Bar a}+\frac{\dot \alpha_2}{2\alpha_2}\Bar a&=& N\Bar  b\\
    \dot {\Bar b}&\approx& -\frac{N\Bar a}{2\alpha_2 } \Bar \rho \\
    \dot {\Bar \rho}+3\frac{\dot{\Bar  a}}{\Bar a}\Bar\rho&=&0
\end{eqnarray}
resulting from the approximate Hamiltonian constraint (see Section~\ref{HamStat}). For the global variables we have:
\begin{eqnarray}
    \dot T_R&\approx &
  \frac{N}{\alpha_2} {\Bar{a}^3}\frac{\Bar\rho}{2}  
    \\
    \dot \alpha_2&\approx &
    - N\alpha_2  \frac{d c_g^2}{d T_R}
   \Bar{a}\Bar{k} .
\end{eqnarray}
As explained in Section~\ref{HamMach} (first comment at the end), in a Machian setting the global variables are an external input into the local physics and their solutions can be copied over from their MSS counterparts.

\begin{figure}[ht]
\includegraphics[scale=0.8]{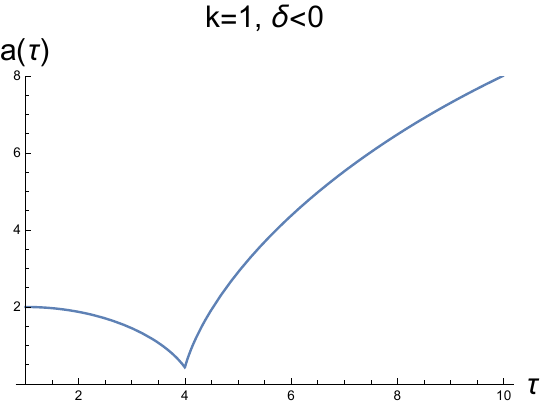}
\caption{The effects of a negative pulse in $c_g^2$ on a standard OS model (with $k=1)$. As we see the collapse is reversed by the pulse. The bounce happens just before the dust ball falls through the Schwarzchild radius because, that is where a pulse in the far future is felt.} \label{ClosedCol}
\end{figure}

\begin{figure}[ht]
\includegraphics[scale=0.8]{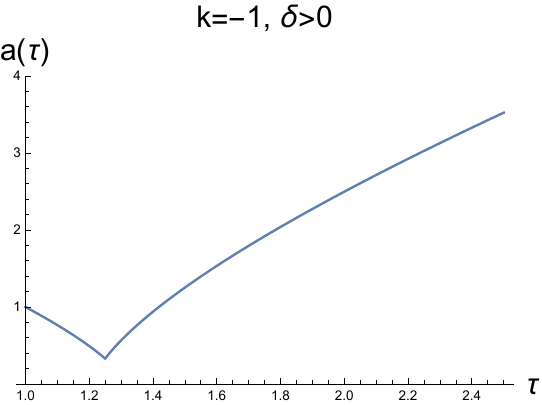}
\caption{A similar bounce obtained with a $k=-1$ version of the OS model and a positive pulse in $c_g^2$. In this case there is no chance of a recollapse. The black hole never forms and the collapsing star is devolved as an explosion into the far future.} \label{OpenCol}
\end{figure}

Specifically, for simplicity, we could choose $\bar k=0$ so that $\dot\alpha_2=0$, leading to a simplified (\ref{dotaG}). We then solve (\ref{dotaG}) and (\ref{dotbG}) assuming 
a pulse in $c_g^2=\delta_{c_g^2}\, \delta(\tau-\tau_\star)$. In the $N=1$ gauge they can be combined into:
\begin{equation}
    \ddot a +  \frac{\dot a^2}{2a}+\frac{k c_g^2}{2a}=0.
\end{equation}
and solved numerically. Alternatively, Eqs.~(\ref{dotaG}) and (\ref{dotbG}) imply a step in $b$ and a cusp in $a=a_\star$,  with:
\begin{equation}
    \Delta b=-\frac{k\delta_{c_g^2}}{2a_\star}.
\end{equation}
By either method, we can easily find scenarios where the collapse is reversed. In Figs.~\ref{ClosedCol} and~\ref{OpenCol} we consider both the standard OS model $k=1$ with a pulse with $\delta_{c_g^2}<0$ (so that the gravitational geometry goes briefly euclidean), and an alternative OS model with $k=-1$ (as described in~\cite{Stephani,Guth-OS}) and $\delta_{c_g^2}>0$. 

Thus we see that a collapsing star, under the effects of evolution to the far future, could actually explode into the far future just as it was about to cross its Schwarzchild radius.
But as these examples show, this feature is not generic within the space of theories in Section~\ref{OS-evol}.

\section{Mass top-up and deficit}\label{top-up}
Whether a collapsing star forms a black hole or explodes, its mass is generically changed by evolution in an energy transfer process between the Machian and local Universe. In contrast with the bounce phenomenon, this process is generic (apart from very fine-tuned exceptions) and is studied in detail in~\cite{evol-CDM}, but can be easily seen here. 

Consider first the cases of the last Section. The fact that evolution is confined to a short period around
$\tau=\tau_\star$ does not mean that the local Hamiltonian constraint (assumed valid for $\tau<\tau_\star$) holds true for $\tau>\tau_\star$. The arguments in Section~\ref{HamStat} only imply that $\dot {\cal H}_L=0$ once evolution stops. Therefore  ${\cal H}_L$ must be a constant at $\tau>\tau_\star$, but this constant generically suffers a step at $\tau_\star$ given by:
\begin{equation}
    \Delta {\cal H}_L=\int dt\, \dot  {\cal H}_L=
    \int dt\, \{ {\cal H}_L,H \}_{NL}
\end{equation}
where we can now use the various expressions found in Section~\ref{HamStat}. In our case this can be more easily read off as:
\begin{equation}\label{DelHpulse}
    \Delta {\cal H}_L=- a_\star \alpha_2 \Delta b^2=
    - a_\star \alpha_2 (2b_{-} \Delta b +(\Delta b)^2)
\end{equation}
where $b_-=b(\tau_{\star -})$, and we have used (\ref{HL}). As comparison with (\ref{HL}) promptly reveals, this is equivalent 
to preserving ${\cal H}_L=0$ and adding to it a {\it new} pressureless component ($w_H=0$) with constant of motion $m_H=-\Delta {\cal H}_L$ and energy density
\begin{equation}
 \rho_H=\frac{m_H}{a^3} .  
\end{equation}
Since $\chi_0$ has not changed the total mass of the star changes. The only exception is when $\Delta b=-2 b_-$, that is when the $b_+=-b_-$ (a perfectly symmetric bounce), but this requires fine-tuning $\delta_{c_g^2}$, and this is an arbitrary external input. As (\ref{DelHpulse}) shows, we can have a top-up or a deficit during a bounce in this model depending on whether $\Delta b>2|b_-|$ or $|b_-|<\Delta b<2|b_-|$, respectively. 

We can consider any scenarios following from the theories in Section~\ref{OS-evol}, with or without bounces, to find this phenomenon, but such an exercise would be dull and pointless to present here.

\section{Conclusions}

In this paper we studied the effect of non-local (or ``global'') physics on the formation of black holes. As perusal of our calculations shows, our findings are generic to theories with non-local, Machian structure, but the specific setting we used was that of {\it evolution } in the laws of physics. Evolution begs the question: in terms of what time? A possible answer is provided by global time variables, dual to the constants of Nature, as defined by a procedure mimicking the Henneax-Teitelboim formulation of unimodular gravity~\cite{unimod}. Such times are foliation-dependent, a matter which has no physical effects, unless we use them to define evolution~\cite{evolution}. In the latter case, the preferred foliation induces a global structure leading to non-local interactions. 

It was further argued in~\cite{evolution} that evolution, and its concomitant energy production, does not need to be confined to the early Universe. It could be happening here and now with subtle effects on a cosmological scale, or to our far future, with far more dramatic consequences. Future cataclysms could even mirror those that led to the Big Bang, to create a rough cyclic structure. Given the association of such future events with global leaves $\Sigma_t$ as $t\rightarrow\infty$, this would be reflected in the physics near the horizon of a Black hole, and in particular in the last infinitesimal moments of collapse, before a black hole is formed. 

We found that it was possible (but not general) for such global effects to reverse the collapse. For some evolution theories all the collapsing stars will explode into the far future, rather than form black holes. The implications for the information paradox are obvious, assuming the problem is real and not self-inflicted (as are many supposed physics problems of our time). Independently of this matter we found that the mass of the black hole, if formed, or exploding star, otherwise, is changed as a result of the interaction with the whole leaf. While evolution is taking place the Universe is in perfect Spinozian/Machian unity. 

We close by noting that our arguments, using a spherically symmetric collapsing dust ball, are  
just an idealized illustration. It was enough to make our points, but questions arise as to what would happen in more realist settings. What would be the impact of pressure? What would be the impact of a peculiar velocity with respect to the cosmological frame (given that the effect is frame dependent)? More importantly, perhaps, how would the black hole rotation (the ultimate Newton's bucket) affect these considerations? 
We also focused on Einstein-Cartan theory, but some of our statements are general and only depend on the existence of an horizon and future evolution.  Could there be novelties in other ``base'' theories (see e.g.~\cite{paolo})? More generally one may wonder what happens to theorems, such as the no-hair and Birkhoff's theorems, if global interactions are switched on.

\section{Acknowledgments}
We thank Bruno Alexandre and Toby Wiseman for comments, and Farbod Rassouli for assistance with the figures. This work was partly supported by the STFC Consolidated Grants ST/T000791/1 and ST/X00575/1.

\end{document}